# Progress on the Construction of the 100 MeV / 100 kW Electron Linac for the NSC KIPT Neutron Source

CHI Yun-Long(池云龙), PEI Shi-Lun(裴士伦)[1)], PEI Guo-Xi(裴国玺), WANG Shu-Hong(王书鸿), CAO Jian-She(曹建社), HOU Mi(侯汨), LIU Wei-Bin(刘渭滨), ZHOU Zu-Sheng(周祖圣), ZHAO Feng-Li(赵风利), LIU Rong(刘熔), KONG Xiang-Cheng(孔祥成), ZHAO Jing-Xia(赵敬霞), DENG Chang-Dong(邓昌东), SONG Hong(宋洪), LIU Jin-Tong(刘晋通), DAI Xu-Wen(戴旭文), YUE Jun-Hui(岳军会), YANG Qi(杨奇), HE Da-Yong(何大勇), HE Xiang(贺祥), LE Qi(乐琪), LI Xiao-Ping(李小平), WANG Lin(汪林), WANG Xiang-Jian(王湘鉴), MA Hui-Zhou(麻惠洲), ZHAO Xiao-Yan(赵晓岩), SUI Yan-Feng(随艳峰), GUO Hai-Sheng(郭海生), MA Chuang-Xin(马创新), ZHAO Jian-Bing(赵建兵), CHEN Peng(陈鹏)

Institute of High Energy Physics, Chinese Academy of Sciences, Beijing 100049, China

**Abstract:** IHEP, China is constructing a 100 MeV / 100 kW electron Linac for NSC KIPT, Ukraine. This linac will be used as the driver of a neutron source based on a subcritical assembly. In 2012, the injector part of the accelerator was pre-installed as a testing facility in the experimental hall #2 of IHEP. The injector beam and key hardware testing results were met the design goal. Recently, the injector testing facility was disassembled and all of the components for the whole accelerator have been shipped to Ukraine from China by ocean shipping. The installation of the whole machine in KIPT will be started in June, 2013. The construction progress, the design and testing results of the injector beam and key hardware are presented.

**Key words:** electron linac, neutron source, beam testing, key hardware testing

**PACS:** 29.20.Ej, 29.25.Dz, 29.27.Eg

## 1. Introduction

The Kharkov Institute of Physics and Technology of National Science Centre (NSC KIPT, Kharkov, Ukraine) together with Argonne National Laboratory (ANL, USA) develops the project of a neutron source based on the subcritical assembly driven by an electron linac with high average beam power [1].The main purposes of the project are to support the nuclear industry and the medical researches. Reactor physics and material researches will be carried out. The goal is to create in Ukraine the experimental basis for the neutron research based on the safe intensive neutron sources.

Two main parts of the neutron source facility are an electron linac and a beam transport line from the linac to the target, both of which are designed by IHEP, China. The linac should be able to provide 100 MeV beam with average power of 100 kW. The beam line should be able to provide a beam transfer with minimum beam losses and form a homogeneous particle density distribution at the target. Construction of such an accelerator with high average beam power and low beam power losses is a technical challenging task, and all components of the machine have to be designed, fabricated, tested, assembled and commissioned elaborately [2-4].

In 2012, the injector testing facility was pre-installed in the experimental hall #2 of IHEP, and the beam and key hardware were tested with satisfying results obtained. In the meantime, the performance of some hardware was also improved by modifying the initial design. Recently, the injector testing facility was disassembled and all of the components for the whole machine have been shipped to KIPT from China by ocean shipping. In early June of 2013, the machine will be assembled in KIPT by an IHEP and KIPT joint team, hopefully the accelerator conditioning and commissioning will be started soon.

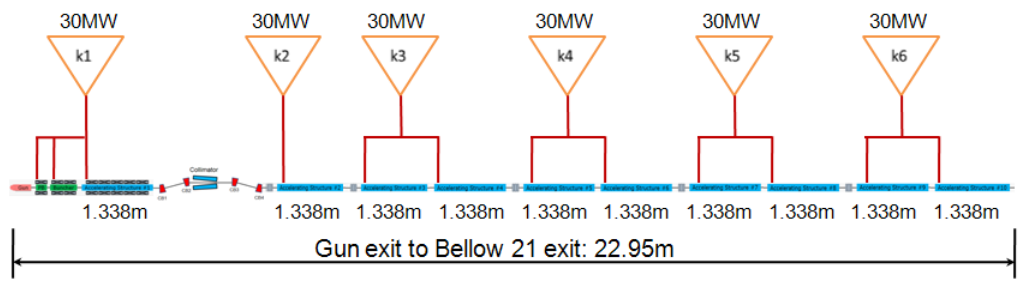

Fig. 1. The schematic layout of the NSC KIPT electron linac.

1) E-mail: peisl@mail.ihep.ac.cn

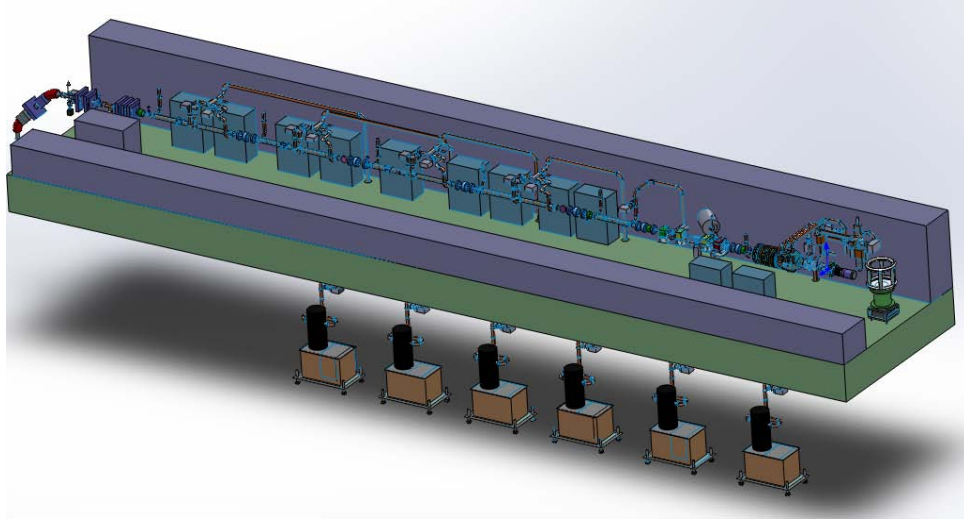

Fig. 2. The mechanical layout of the accelerator tunnel and the klystron source distribution.

## 2. Linac layout and main parameters

Figure 1 shows the schematic layout of the whole linac with main parameters listed in Table 1. Fig. 2 shows the mechanical layout of the accelerator tunnel including the beam transport line to the target and the RF power source distribution. The klystron gallery is located under the accelerator tunnel.

Table 1. Main parameters of the NSC KIPT linac

| parameters | Values |
|---|---|
| RF frequency / MHz | 2856 |
| beam energy / MeV | 100 |
| beam power / kW | 100 |
| beam current (max.) / A | 0.6 |
| energy spread (peak-to-peak) | ±4% |
| emittance / m-rad | $5\times10^{-7}$ |
| beam pulse length / μs | 2.7 |
| RF pulse length / μs | 3 |
| pulse repetition rate / Hz | 625 |
| klystron | 6 units / (30 MW / 50 kW) |
| accelerating structure | 10 units / 1.338 m |
| gun high voltage / kV | ~120 |
| nominal gun beam current | ~1–1.2 A |

To satisfy the peak-to-peak energy spread requirement at the linac exit, particles with large energy difference from the synchronous particle should be eliminated at the low energy stage to ease the design of the beam collimation and the radiation shielding systems. A dispersion free chicane system is introduced and located downstream the injector part but upstream the 2$^{nd}$ accelerating structure. There are 4 bending magnets in the chicane system, mechanical layout of which is shown in Fig. 3. The bending magnets CB1 and CB4 are sectors, while CB2 and CB3 are rectangles. The 1$^{st}$ bending magnet CB1 was specially designed to have two functions. One is for the nominal beam collimation process with a bending angle of 10$^{o}$; another is to be used as an energy analyzing magnet (AM) with 45$^{o}$ bending angle at the injector exit, and the RF phases seen by the electron beam in the injector can be optimized in the real operation. The vacuum chambers for CB1, AM and CB2 are integrated together for accelerator longitudinal installation space saving reason. Similarly, the chamber for CB3 and CB4 are also an integrated one.

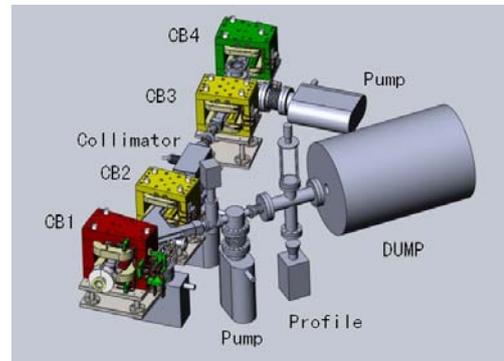

Fig. 3. The layout of the chicane system.

## 3. Injector testing and upgrade

Figure 4 shows the schematic layout of the linac injector. To get a clean bunch without any satellite electrons in each RF bucket downstream the chicane system, the phases of all the RF structures and the solenoid field along the injector are tuned to obtain the optimized beam phase and energy spectrums shown in Fig. 5 at the injector exit, they are appropriate for the downstream beam collimation.

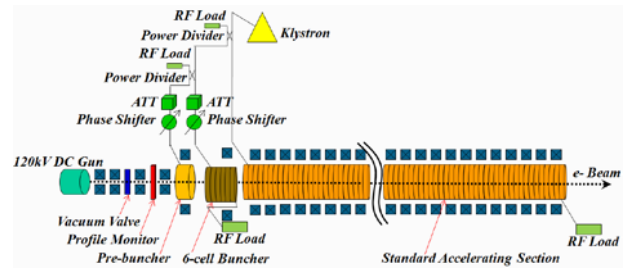

Fig. 4. The schematic layout of the linac injector.

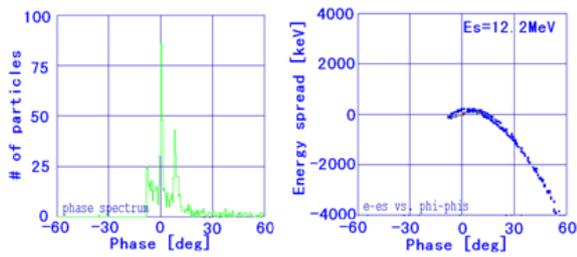

Fig. 5. The optimized beam phase and energy spectrums at the injector exit.

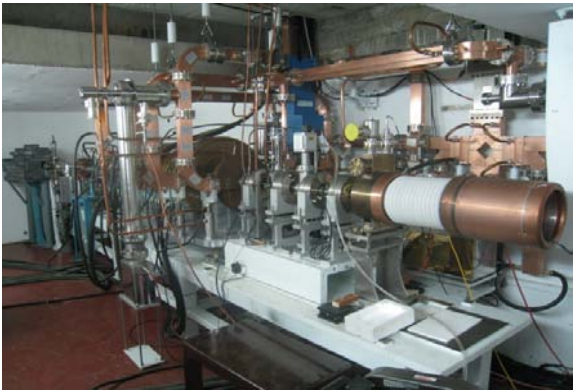

Fig. 6: The injector testing facility installed in IHEP.

Figure 6 shows the injector testing facility installed in the experimental hall #2 of IHEP in 2012. Initial beam testing showed that the beam current signals measured by BCT and ACCT couldn't reflect the true beam pulse shape because of the relatively longer rise/fall time [4]. In the meantime, due to the interference of the grounding system, the beam current waveforms' swung up and down along the whole beam pulse, the beam tuning and bunching efficiency estimation could only be done very roughly. Finally, by replacing the BCT and ACCT with FCT, improving the grounding system and the beam tuning, the beam transportation efficiency in the injector exit was increased to ~90%. Fig. 7 shows the measured beam current signals with the nominal value of ~1.1 A/2.7μs at the electron gun exit. FCT1 and FCT2 are located at the exits of the gun and injector, respectively. The maximum beam current obtained at the injector exit is ~2 A with ~3 μs pulse length, this is limited by the electron gun capability, no clear BBU effect observed in the testing facility for this scenario, thus it is believed that ~600 mA beam can be successfully obtained at the downstream main linac part.

It is worth to point out that the 2.7 μs pulse length is not so critical here; the ultimate goal is to obtain 100 MeV/100 kW electron beam. A lot of measures have been adopted during the machine design stage to assure the beam performance of a 2.7 μs/600 mA beam at the linac exit [2]. However, there is still possibility that only <2.7 μs / <600 mA beam can be operated in the real machine operation due to the regenerative beam break-up (BBU) effect, the solution for this scenario is to shorten the beam pulse length but increase the machine repetition frequency to be higher than 625 Hz.

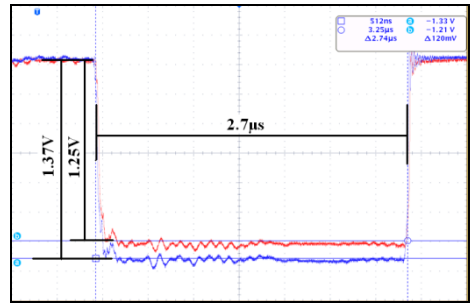

Fig. 7. The measured FCT1 (blue) and FCT2 (red) signals.

By measuring the beam profile following a dipole analyzing magnet located downstream of the injector, the 1σ beam energy spread at the injector exit were calculated to be ~2%. Fig. 8 shows one typical beam profile measured at the energy and energy spread measurement line. During the whole measurement process, the injector beam transportation efficiency was stabilized to be ~90%, and no clear high energy beam tails were found but only the low energy tails, which means electron beams provided by the injector are very appropriate for the beam collimation process with the following chicane system to eliminate all particles with very large beam energy and phase spreads. By this way, the beam power losses along the beam transport line can be well minimized.

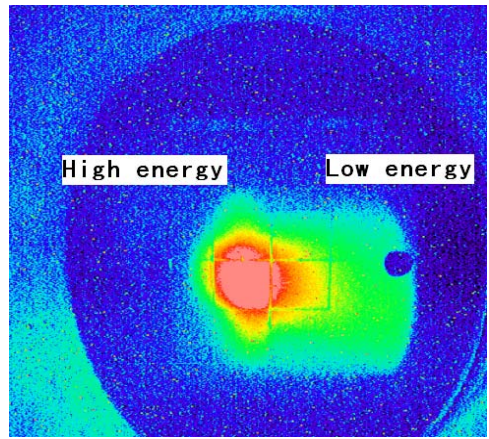

Fig. 8. One typical beam profile measured at the energy and energy spread measurement line.

By optimizing the klystron drive waveform and applying feed-forward and feedback techniques, the beam loading compensation system was also tested. It is found that the injector's beam energy spread can be further decreased a little bit, which is expected and validates its functionality.

## 4. Main systems and components

The constructions of all the main accelerator systems have been completed in early March, 2013. Later, all the auxiliary components are prepared and tested.

### 4.1 Electron gun

The thermionic electron gun with Y824 cathode assembly has been in steady turn-key operation during the whole injector testing period. The testing shows that >2

A/3.0 μs/120 keV beam can be produced stably [5]. Maximum high voltage (HV) of 150 kV can be provided by the HV power station shown in Fig. 9.

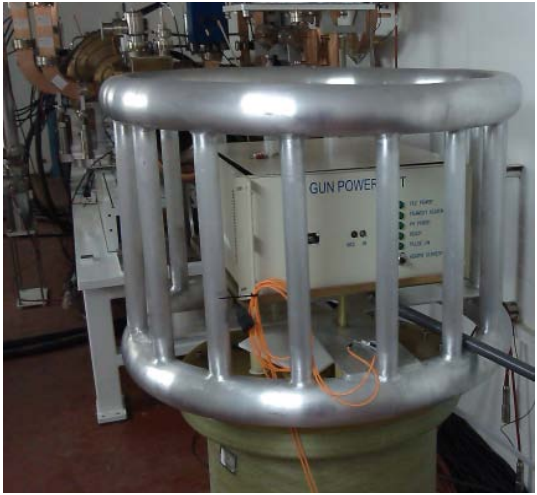

Fig. 9. The high voltage station for the 120 kV gun.

### 4.2 RF structures

The pre-buncher is a single cell standing wave (SW) cavity as shown in Fig. 10. The RF power is fed into the cavity by rectangle waveguide with measured coupling factor β=~1.73. The cavity resonates at 2856 MHz with a bandwidth of ±3.2 MHz.

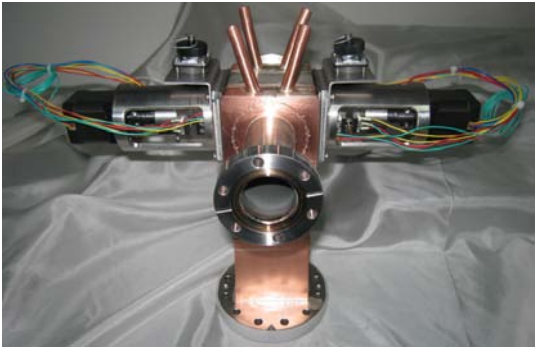

Fig. 10. The pre-buncher.

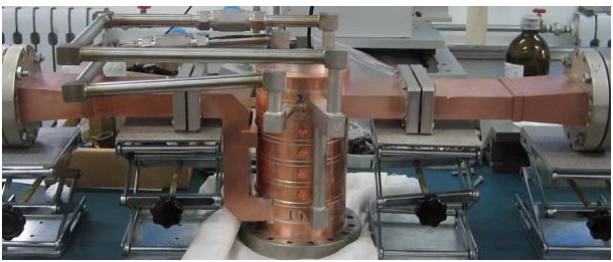

Fig. 11. The 6-cell travelling wave buncher.

The buncher is a travelling wave (TW) constant impedance (CI) structure with phase velocity of β=0.75. Initially, the buncher was designed to have only 4 cells. However, the adoption of water cooling jacket demands more longitudinal space (leading a longer buncher) to ease the installation. Finally, one 6-cell version was developed, which is shown in Fig. 11. The measured VSWR at 2856 MHz is ~1.02 with a bandwidth of ~5.5 MHz (VSWR ≤ 1.2). The measured filling time and attenuation factor are ~50 ns and ~0.56 dB, respectively.

Ten TW constant gradient (CG) accelerating structures (A0–A9 along the Linac) with relatively bigger beam aperture have been developed and will be installed in the KIPT accelerator tunnel to boost the beam energy to 100 MeV. To suppress the BBU (both regenerative and cumulative) effect, ~1.3 m long 2π/3 mode quasi-constant gradient structure was adopted. The iris aperture decreases from 27.887 mm to 23.726 mm in a stepwise fashion along the structure (26.220 mm to 19.093 mm for BEPCII 3 m long structure). To detune the dipole mode, its frequency spread was increased by increasing the disk hole diameter step to ~0.122 mm (~0.085 mm for the BEPCII 3 m long structure). At the $2^{nd}$ to $6^{th}$ disks of each structure from A1–A9, 4 holes with diameters of 9 mm (A1, A4 and A7), 11 mm (A2, A5 and A8) and 13 mm (A3, A6 and A9) were drilled. By this way, the HEM11 mode frequency will be increased a certain amount in these cells. Detail specifications of the ~1.3 m long structure can be found in Reference [6].

Fig. 12 shows one structure in the RF cold testing lab. All of the 10 structures were tuned to have a cell-to-cell phase error ≤±0.5° and a cumulative one to the $1^{st}$ cell ≤±2°. The measured bandwidth, the attenuation factor and the filling time are ~5.5 MHz (VSWR≤1.2), ~0.155 Np and ~220 ns, respectively, which are consistent with the RF design [6]. Water cooling jacket is mounted along each structure after the RF tuning in the lab and will be used to cool down the structure during the machine operation with high average RF input power.

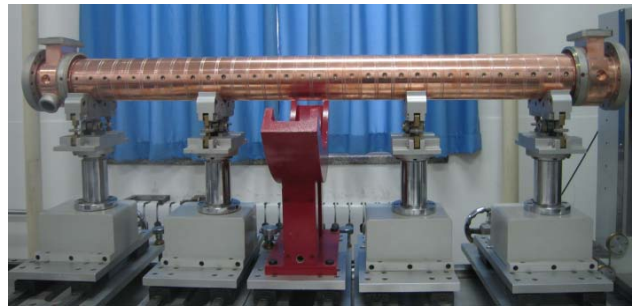

Fig. 12. The accelerating structure in RF cold testing lab.

### 4.3 RF source

Six RF power units with each consisted by a Toshiba E37311 klystron shown in Fig. 13 and its modulator made in China are applied in the KIPT linac.

The $1^{st}$ RF power unit shown in Fig. 14 used in the injector testing facility to energize the pre-buncher, the buncher and the $1^{st}$ accelerating structure has been conditioned up to 500 Hz repetition rate with 3.2 μs pulse width (flat-top), and limited by the electrical capability of the experimental hall #2. Fig. 15 shows the corresponding klystron output waveforms with 27 MW peak output power at 500 Hz repetition rate. Fig. 16 shows the klystron output power at different modulator high voltages.

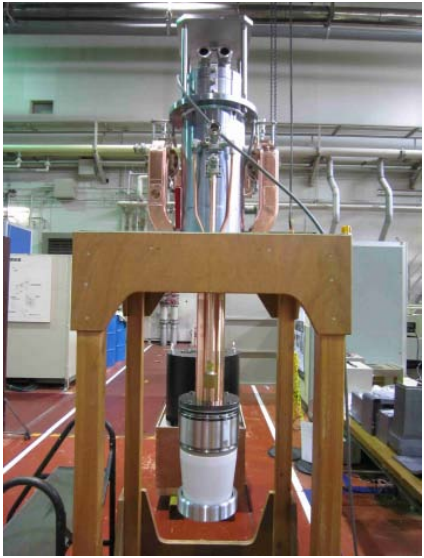

Fig. 13. The Toshiba E37311 klystron.

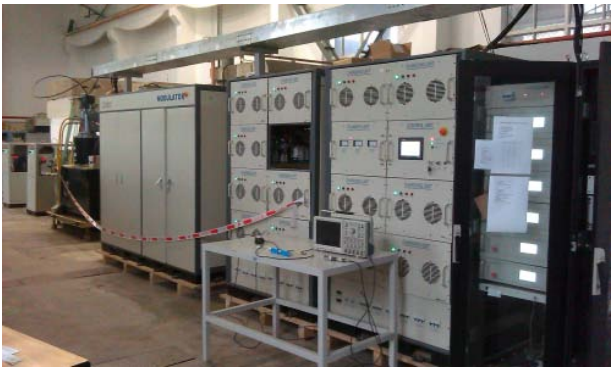

Fig. 14. The modulator made in China.

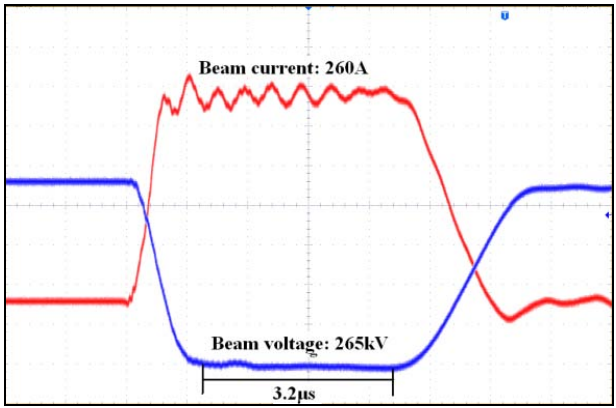

Fig. 15. The klystron output waveforms with 27 MW peak output power at 500 Hz repetition rate.

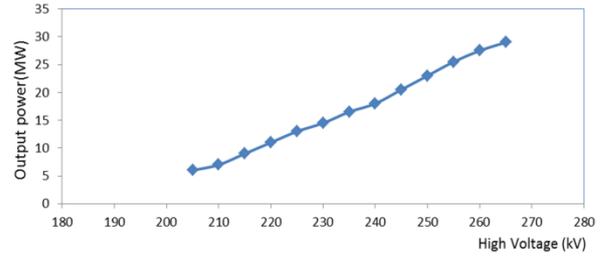

Fig. 16. The relationship between the klystron output power and the modulator high voltage.

### 4.4 Beam instrumentation

The beam instrumentation system is capable of measuring the beam positions, the beam intensities, the beam profile, the emittance, the beam loss, the beam energy and the energy spread, etc. These parameters and information are very important for the machine commissioning and operation.

Fig. 17 shows the beam instrumentation devices distribution along the linac and the transportation line to the target. In the KIPT linac, 8 button type BPMs and BLMs are used to measure the beam orbit and beam losses; 3 PRs and 2 WSs are used to measure the beam profile; 5 FCTs with very fast rise/fall time are used to measure the beam current and pulse shape.

There are total two beam energy analyzing stations located at the exits of the injector and the linac, respectively. The Strip Lame Screen (SLS) is used to measure the beam profile downstream the analyzing magnet, by which the beam energy and energy spread can be determined. Fig. 18 shows the SLS assembly, which can withstand relatively higher average beam power, thus the beam energy can be measured more accurately with a relatively beam higher repetition rate. Because of the uncertainty of the beam TWISS parameters at the linac exit, the quadruple scanning method by a triplet is used to measure the beam emittance and the TWISS parameters there, which is very helpful for the transport line tuning.

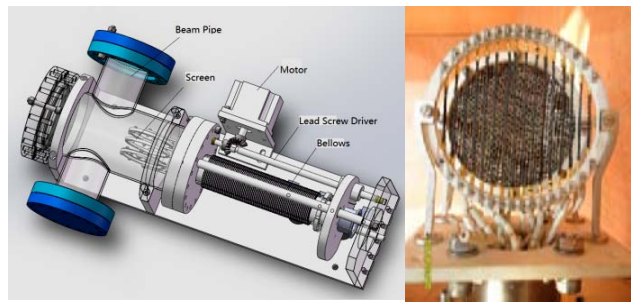

Fig. 18. The Strip Lame Screen assembly.

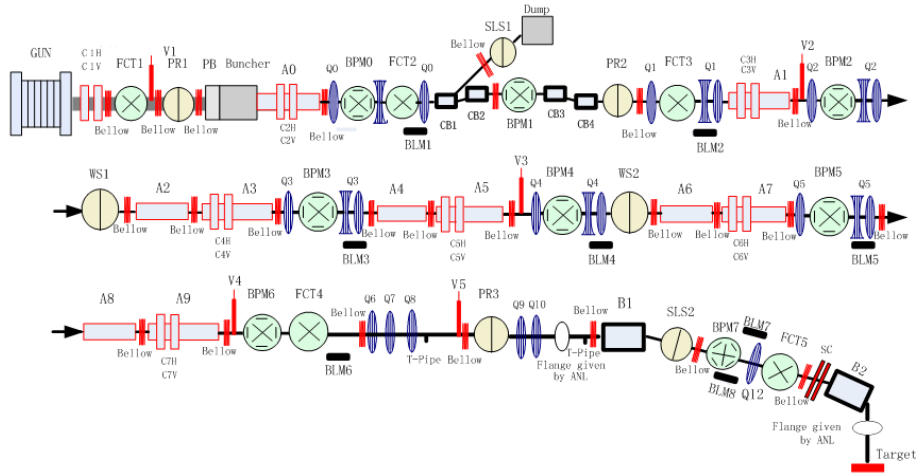

Fig. 17. The beam instrumentation devices distribution along the linac and the transportation line to the target.

### 4.5 LLRF system

The LLRF system consists of 6 control units, which are able to control and adjust the RF field amplitude and phase, generate drive waveform for the RF amplifier, monitor the whole RF system, realize beam loading compensation and data analysing, etc. The drive signal of each unit comes from the reference signal distributor. Feed forward is used to set up the drive waveform, while feedback for optimization of the drive phase, which comes from the beam phase measurement cavity. Initial testing shows the desired RF field shape and the optimized beam phase in the accelerating structure can be obtained. Fig. 19 shows one typical control unit.

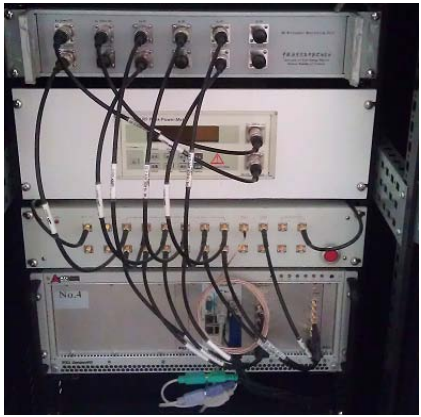

Fig. 19. One typical control unit.

### 4.6 Control system

The control system is EPICs based with Channel Access communication protocol. The interactive interface is developed with Control System Studio (CSS). Online database is realized by Channel Archiver. Most of the components have been tested in the injector testing facility. Fig. 20 shows the newly updated control system architecture.

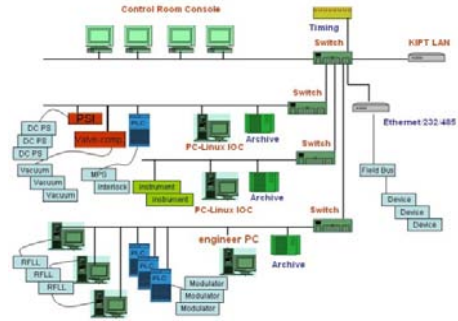

Fig. 20. The newly updated control system architecture.

### 4.7 Magnet system and beam transportation line

The whole machine magnet system consist of 4 gun focusing lens, 22 solenoids, 6 triplets, 7 correctors and 4 chicane dipoles. The $1^{st}$ chicane dipole was specially designed to have two functions [4]. One is for the nominal beam collimation process; another is to be used as an energy analyzing magnet (AM). For the beam transport line with schematic layout shown in Fig. 21, there are 6 quadruples, 2 dipoles with $45^{o}$ bending angle and 1 pair of scanning magnets (horizontal and vertical). The Q11 is used to cancel the dispersion.

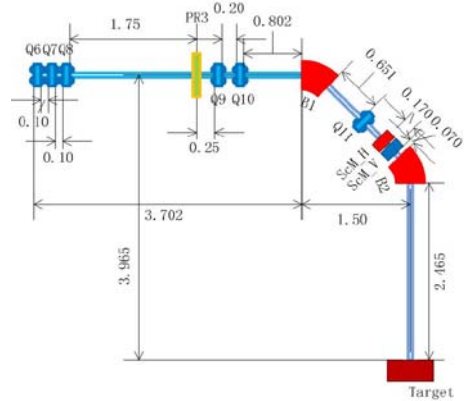

Fig. 21. The beam transportation line.

By using the scanning magnets, a homogenous beam intensity on the target shown in Fig. 22 can be formed in one scanning period. Theoretically, the beam density uniformity can reach ~5% at the target.

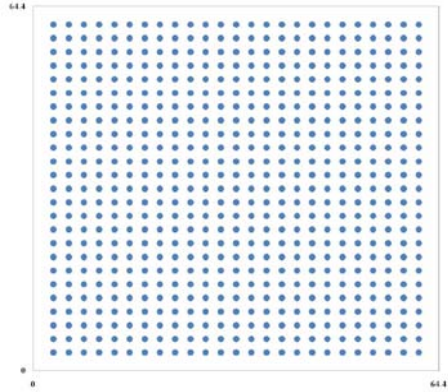

Fig. 22. The homogenous beam distribution on the target.

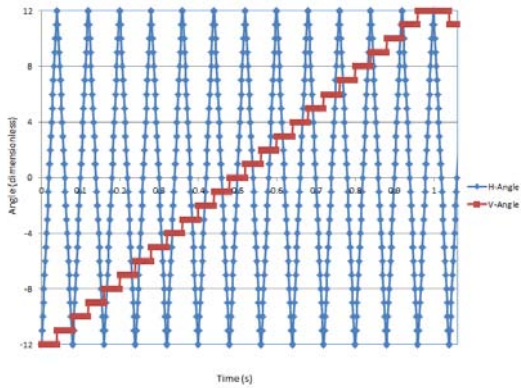

Fig. 23. The scanning magnets' strength setting in 1 s.

According to a repetition rate of 625 Hz, both horizontal and vertical scanning magnets need to run 25 steps in 1 second. Because the beam pulse time interval is very short (1.6 ms), the switching frequency of one magnet should be 12.5 Hz with saw tooth waveform (blue line in Fig. 23). Another magnet strength switches with multi-step (red line in Fig. 23), and the step is very steep (1.6 ms).

Fig. 24 shows the 1$^{st}$ chicane dipole and one of the scanning magnets in field measurement. The magnetic field of all magnets has been measured and meets the design requirement.

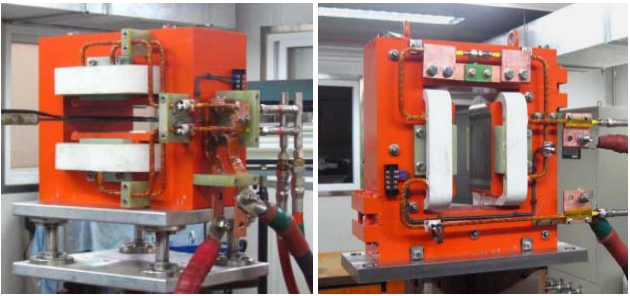

Fig. 24: The 1$^{st}$ chicane dipole (left) and one of the scanning magnet (right).

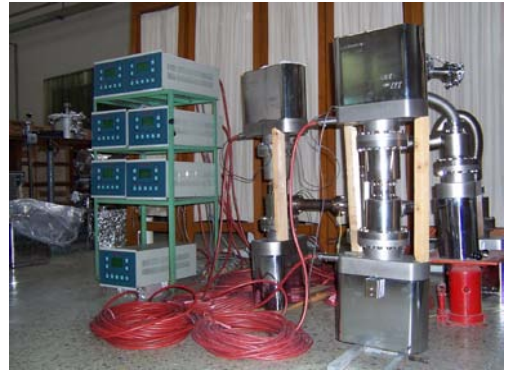

Fig. 26. The ion pumps testing system.

### 4.8 Vacuum system

The vacuum system shown in Fig. 25 divides the whole machine to 7 sections by 6 gate valves, which assures the accelerator working at ultra-high vacuum condition—better than $5.0 \times 10^{-8}$ mbar. All ion pumps are made in China and have been off-line tested by an ion pumps testing system shown in Fig. 26.

On-line testing of the vacuum system in the injector testing facility shows the vacuum system can meet the design goal.

### 4.9 Water cooling system

The water cooling system is designed to be composed by 3 subsystems: 1) first loop of 30 (max.)±1$^{o}$C for the tunnel devices; 2) first loop of 30 (max.)±1$^{o}$C for the klystrons gallery devices; 3) 40±0.2$^{o}$C constant temperature system for the accelerator.

The water cooling system will be the 1$^{st}$ sub-system installed in the KIPT Linac. The prototype has been successfully tested in the injector testing facility.

## SUMMARY

The construction of the 100 MeV / 100 kW electron linac by IHEP, China for the NSC KIPT neutron source is going on smoothly. In 2012, the injector part of the linac was pre-installed as a testing facility in IHEP. Almost all of the main systems and components were tested in the injector testing facility with satisfying results. The construction of all the accelerator systems have been completed in early March, 2013. Later, all the other auxiliary components are prepared and tested. Recently, the injector testing facility was disassembled and all of the components for the whole linac have been shipped to KIPT.

In early June of 2013, the machine will be assembled in KIPT by an IHEP and KIPT joint team, hopefully the accelerator conditioning and commissioning will be started soon.

*The authors would like to thank their KIPT and ANL colleagues, especially A. Zelinsky and M. Ayzatskiy of KIPT and Y. Gohar of ANL, for their very strong support and helpful discussions.*

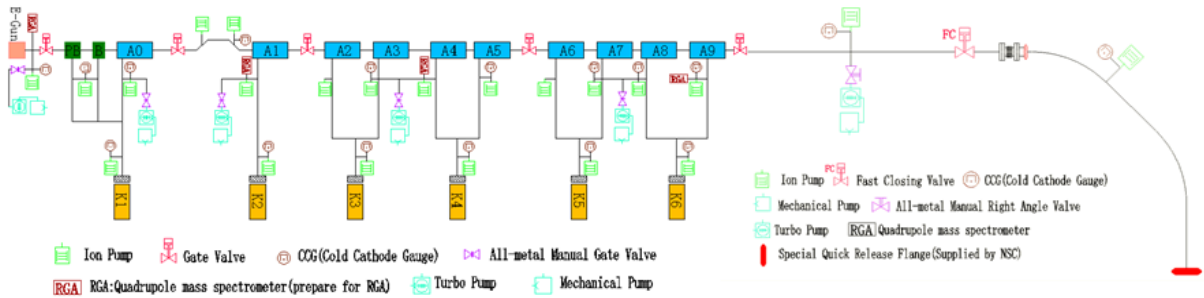

Fig. 25. The vacuum system layout.